# Review and Prospect: Deep Learning in Nuclear Magnetic Resonance Spectroscopy

Dicheng Chen[#][a], Zi Wang[#][a], Di Guo[b], Vladislav Orekhov[c], Xiaobo Qu*[a]

**Abstract:** Since the concept of Deep Learning (DL) was formally proposed in 2006, it had a major impact on academic research and industry. Nowadays, DL provides an unprecedented way to analyze and process data with demonstrated great results in computer vision, medical imaging, natural language processing, etc. In this Minireview, we summarize applications of DL in Nuclear Magnetic Resonance (NMR) spectroscopy and outline a perspective for DL as entirely new approaches that are likely to transform NMR spectroscopy into a much more efficient and powerful technique in chemistry and life science.

## 1. Introduction

With the rapid development of experimental techniques, Nuclear Magnetic Resonance (NMR) spectroscopy finds new applications in chemistry, life sciences, and other fields. It provides atomic-level information on molecular structure and is an indispensable tool for the analysis of molecular dynamics and interactions.

Although early demonstrations of machine learning in NMR spectroscopy appeared in the 1970s [1], practical applications had to await the next generation of algorithms and modern computing power. In recent years, artificial intelligence technology attracted great interest in various research fields because of the availability of high-performance hardware like Graphics Processing Units (GPUs). Deep Learning (DL) is a representative artificial intelligence technique utilizing neural networks. DL can discover essential features embedded in large data sets and figure out complex nonlinear mappings between inputs and outputs [2], and thus does not require any prior knowledge or formal assumptions (Figure 1). To date, DL has been successfully demonstrated in different areas, including computer vision [2], medical imaging [3], NMR spectra reconstruction [4], magnetic resonance spectroscopic imaging [48] and biological data analysis [5]. In view of the clear success, more and more researchers in the NMR field start to pay attention to DL and explore it for addressing deficiencies of conventional methods.

The following discussions will focus on four common practical problems. Firstly, NMR data acquisition by undersampling of the regular Nyquist grid is the most direct and important method for reducing the measurement time. Inevitably, a spectra reconstruction procedure is needed to repair the information loss caused by undersampling, and DL represents a powerful alternative to the existing methods. Secondly, the spectra that we get from spectrometer often have low signal-to-noise ratio (SNR), and thus may benefit from denoising. With the help of a well-trained neural network, the spectra with many interfering signals can be cleaned to increase practical SNR. Finally, DL can also improve the interpretation of the spectra by chemical shift prediction and automated peak picking. We start by introducing the basic DL architectures and the network training process that had been utilized in NMR spectroscopy.

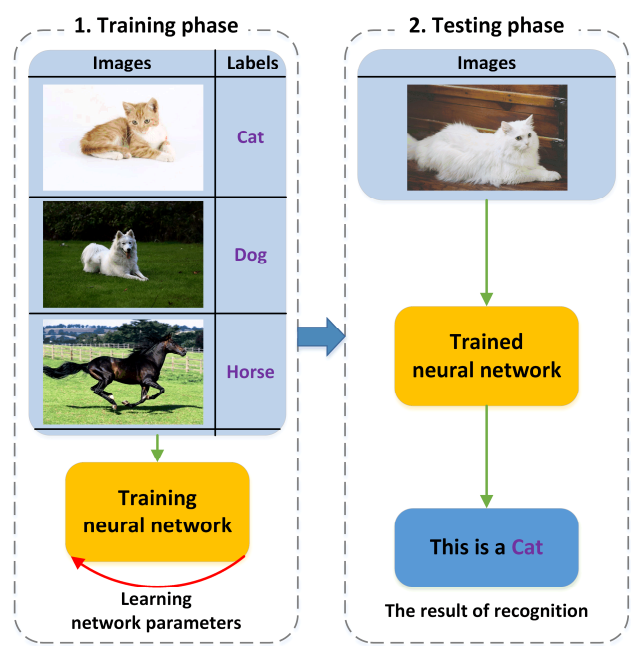

**Figure 1.** A toy example of image recognition with Deep Learning (DL). In the training phase, images and labels of different animals are provided by users. The backpropagation algorithm is then used to adjust the internal parameters of the neural network in such a way that the network learns how to identify animals. In the testing phase, the trained network can correctly recognize animals.

## 2. Basic Architectures of Deep Learning

DL architectures are neural networks consisting of multiple-nonlinear layers. Up to now, DL applications in NMR spectroscopy are mainly based on the following three basic architectures: Deep Neural Networks (DNNs) [6], Convolutional

[a]   D. Chen[#], Z. Wang[#], Prof. X. Qu[*]
      Department of Electronic Science, Fujian Provincial Key Laboratory of Plasma and Magnetic Resonance, State Key Laboratory of Physical Chemistry of Solid Surfaces, Xiamen University
      P.O.Box 979, Xiamen 361005 (China)
      *E-mail: quxiaobo@xmu.edu.cn
[b]   Prof. D. Guo
      School of Computer and Information Engineering, Xiamen University of Technology, Xiamen 361024, China
[c]   Prof. V. Orekhov
      Department of Chemistry and Molecular Biology, University of Gothenburg, Box 465, Gothenburg 40530, Sweden
[#]   These authors contributed equally to this work.

Neural Networks (CNNs) [7] and Recurrent Neural Networks (RNNs) [8]. It is important to note that in this review, we use 'DNNs' to refer mainly to Multi-Layer Perceptron (MLP), which represents fully-connected adjacent networks without convolution units or time association.

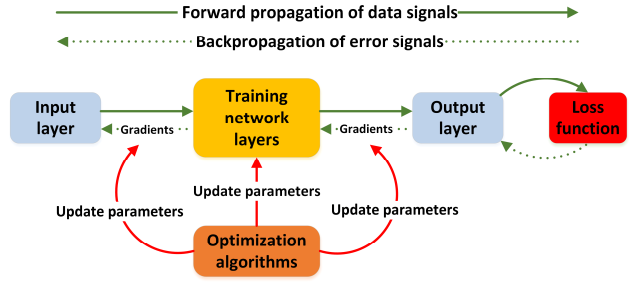

**Figure 2.** The flowchart of neural network training.

The main objective of the network training is to optimize the internal network parameters for each layer (Figure 2). A single step of the optimization process can be briefly summarized as follows. Firstly, given a training data set, the forward propagation computes the output of each layer in sequence and propagates it forward through the network. The loss function measures the error between the inference outputs and the labels. To minimize the loss, the backpropagation uses the chain rule to backpropagate error information and compute gradients of all parameters in the network [9]. Finally, all parameters are updated using optimization algorithms, such as Stochastic Gradient Descent (SGD) [10] or Adam [11]. Besides, regularization, e.g. dropout [12] or Batch Normalization (BN) [13], plays a key role in avoiding overfitting between inference outputs and given labels, and achieving high generalization performance.

Below, we explain each architecture in more detail.

**2.1. Deep Neural Networks**

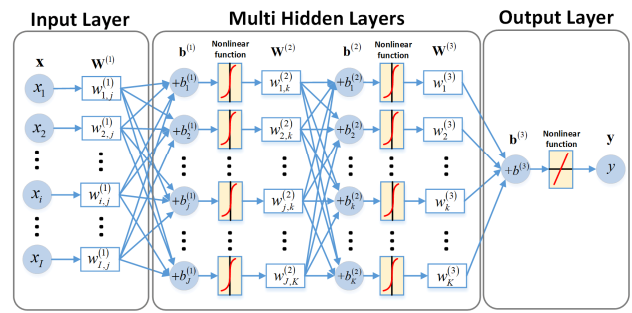

**Figure 3.** The classical structure of DNNs is composed of an input layer, multiple hidden layers, and an output layer. $\mathbf{x} = [x_1, \cdots, x_I]$ is the input vector, $\mathbf{y}$ is the output vector, $\mathbf{W}^{(n)}, \mathbf{b}^{(n)}$ are the weight matrix and the bias vector of the $n^{th}$ full connection.

DNNs are fully-connected, which means that each neuron in every layer is connected to all neurons in the next layer and the size of the neuron input is equal to the number of neurons in this layer. The classical structure of DNNs is composed of an input layer, multiple hidden layers, and an output layer (Figure 3). When the data enter the input layer, the output values are computed layer by layer in the network. In each hidden layer, after receiving a vector consisting of output values of each neuron in the previous layer, it is multiplied by weights imposed by each neuron in the current layer to obtain the weighted sum. The nonlinear function called the activation function, e.g. sigmoid or Rectified Linear Unit (ReLU) [14], is then applied to the weighted sum. Due to these nonlinear functions, the neural network can fit complex nonlinear mappings between inputs and outputs. After passing through all hidden layers, the result is obtained in the output layer.

The forward propagation of DNNs follows the chain rule and can be expressed as follows:
$$\mathbf{y} = f\big(\mathbf{W}^{(n)} f\big(\mathbf{W}^{(n-1)} \cdots f\big(\mathbf{W}^{(1)}\mathbf{x} + \mathbf{b}^{(1)}\big) + \mathbf{b}^{(n-1)}\big) + \mathbf{b}^{(n)}\big), \quad (1)$$
where $\mathbf{x}$ is the input vector, $\mathbf{W}^{(n)}, \mathbf{b}^{(n)}$ are the weight matrix and the bias vector of the $n^{th}$ full connection, $f(\cdot)$ is the activation function and $\mathbf{y}$ is the output vector.

DNNs are especially suited for complex high-dimensional data analysis, not only for the extraction of features but also for the mapping. Given the complexity and high-dimensional nature of NMR spectral data, in the future, DNNs may be more utilized for analyzing complex NMR spectra.

**2.2. Convolutional Neural Networks**

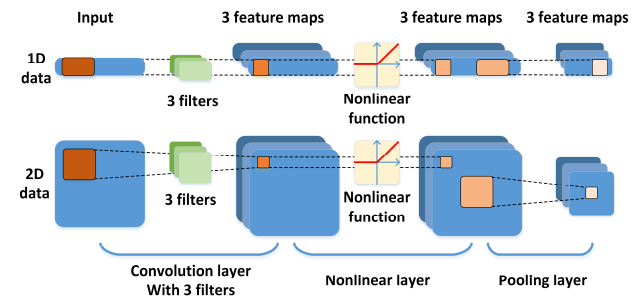

**Figure 4.** The basic structure of CNNs takes 1D and 2D inputs as example. Generally, a CNN is composed of convolution layers, nonlinear layers, and pooling layers.

CNNs are designed to process data from multiple arrays: 1D for sequences, 2D for images and 3D for videos. They are adopted to imitate three key ideas of the brain visual cortex: local connectivity, location invariance, and local transition invariance [2].

Unlike DNNs, CNNs are not directly linked between layers, they use intermediaries called filters which represent weights and biases. Generally, the basic structure of CNNs consists of convolution layers, nonlinear layers, and pooling layers (Figure 4). Each convolution layer obtains groups of local weighted sums, called feature maps, by computing the convolutions (inner products) between local patches of the input maps and the filters. All units in a feature map share the same filter, i.e. same weights and biases, in order to reduce the number of learning parameters. Similar to DNNs, then feature maps pass through nonlinear layers that usually use the ReLU function [15]. The role of pooling layers is to aggregate semantically similar features to identify complex features by making maximum or average subsamples in feature maps. Sometimes, pooling layers are also used to avoid network overfitting and improve the generalization of the model.

Through a convolution layer with $J$ filters, a nonlinear layer and a pooling layer sequentially, we can obtain $J$ output maps:
$$\mathbf{Y}_j = pool\big(f(conv(\mathbf{X}, \mathbf{F}_j) + \mathbf{B}_j)\big), \quad (2)$$
where $\mathbf{X}$ is the input map, $\mathbf{F}_j, \mathbf{B}_j, \mathbf{Y}_j$ ($j = 1,2,\cdots,J$) are the $i\_th$ filter, biases and output map respectively. $conv(\cdot)$ is the convolution operator, $f(\cdot)$ is the nonlinear function and $pool(\cdot)$ is the pooling operator.

Given the excellent ability of CNNs to analyze spatial information, they can be applied to NMR spectra reconstruction, denoising, and chemical shift prediction.

## 2.3. Recurrent Neural Networks

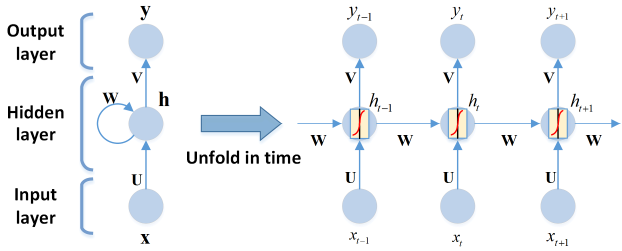

**Figure 5.** The basic structure of RNNs consists of an input unit, a hidden unit, and an output unit with a cyclic connection. The RNNs can be unfolded in time to show the recurrent computation explicitly. $x_t, h_t, y_t$ are the input unit, the hidden unit and the output unit value at $t$, respectively. $\mathbf{U}, \mathbf{V}, \mathbf{W}$ are the weight matrices between different neurons.

For tasks that require processing of sequential inputs, such as time-domain signals, RNNs are often used, which basic structure consists of an input unit, a hidden unit, and an output unit with a cyclic connection (Figure 5). In RNNs, the output of neurons at the current moment directly acts on itself at the next moment, while the data processing of such sequential data by DNNs and CNNs is independent for each moment.

RNNs process one element of the input sequence at a time, store a state vector in a hidden unit that contains information about all previous elements of the sequence, and the current output of the unit needs to take into account both the state vector and the current input into consideration. This property is like a Markov chain of order $n$. If an RNN is unfolded in time (Figure 5), it is even deeper than DNNs and CNNs.

In the forward propagation of RNNs, we can obtain the output $y_t$ at time $t$:

$$y_t = g(\mathbf{V}f(\mathbf{U}x_t + \mathbf{W}h_{t-1}))$$
$$= g\left(\mathbf{V}f\left(\mathbf{U}x_t + \mathbf{W}f(\mathbf{U}x_{t-1} + \mathbf{W}f(\mathbf{U}x_{t-2} + \cdots))\right)\right), \quad (3)$$

where $x_t, h_t$ are the input and the hidden unit value at $t$, respectively. $\mathbf{U}, \mathbf{V}, \mathbf{W}$ are the weight matrices between different neurons, $f(\cdot), g(\cdot)$ is the activation function in the hidden layer and the output layer, respectively.

However, conventional RNNs turn out to be problematic because of the vanishing gradient situation during the training and difficulty of storing data for very long time series [8a]. To solve the problem, the Long Short-Term Memory (LSTM) [8c] networks that use the special hidden unit, were proposed. The special hidden unit called the memory cell achieves the long-term storage through the switch of gate functions.

Since Free Induction Decay (FID) signals and NMR spectra data are sequential, the success of RNNs in natural language processing will provide useful guidance for processing time-domain NMR data.

## 2.4. Deep Learning Libraries

In order to implement DL into applications, one can use several mainstream libraries including TensorFlow [16], Torch [17], Caffe [18], MATLAB neural network toolbox [19] and so on. There are still no clear leaders, and each library has its own advantages.

Table 1 summarizes the software and hardware bases, network architectures and shared resources for the NMR spectroscopy applications cited in the paper.

**Table 1.** The mentioned applications of DL in NMR spectroscopy and their details.

| Applications | Network Architectures | Training Dataset | DL Libraries | Shared Resources | Ref. |
|---|---|---|---|---|---|
| Reconstruction of the spectra | CNN | $4\times10^4$ synthetic FIDs | TensorFlow | https://github.com/She1don23/ | [4a] |
|  | LSTM | $8\times10^6$ synthetic FIDs | TensorFlow | N/A | [4b] |
| Denoising of the spectra | CNN | $4\times10^4$ simulated spectra | MATLAB neural network toolbox | N/A | [20] |
| Chemical shift prediction | DNN | 580-protein database | C++ | https://spin.niddk.nih.gov/bax/software/SPARTA+/ | [21] |
|  | DNN | BioMagResBank [49], Protein Data Bank [50] | N/A | http://spin.ccic.ohio-state.edu/index.php/ppm | [51] |
|  | DNN | 580-protein database, 9523-protein structure database | C++ | https://spin.niddk.nih.gov/bax/software/TALOS-N/ | [22] |
|  | CNN | 2000 crystal structures from the Cambridge Structural Database [23] | TensorFlow | https://thglab.berkeley.edu/software-anddata/ | [24] |
| Automated peak picking | CNN | In the paper [25] | Dumpling | https://github.com/dumpling-bio/ | [5] |
|  | DNN | $2\times10^6$ simulated spectra | N/A | N/A | [26] |

Note: 'N/A' means it is not mentioned in the reference. The mentioned applications of DL in Ref. [4a], [4b], [20] and [24] use GPU acceleration, and the GPU types are NVIDIA Tesla K40M, NVIDIA GeForce GTX 1080 TI, NVIDIA Titan Xp and NVIDIA Tesla P100, respectively.

## 3. Reconstruction of the Spectra

Since the duration of NMR experiments increases rapidly with spectral resolution and dimensionality, Non-Uniform Sampling approach (NUS) [27] is commonly used for accelerating the acquisition of experimental data. Modern methods used for reconstructing high-quality spectra from NUS data [28] rely on prior knowledge or assumptions. Moreover, the algorithms of all these methods are usually iterative and need lengthy computation time to achieve the goal.

DL learns optimal mapping of undersampled FID input signals to target spectra. It can infer the essential features of training data and therefore does not require prior knowledge or assumption. Furthermore, because the network algorithm is non-iterative, has low-complexity, and allows massive parallelization with GPUs, the reconstruction of high-quality spectra through a trained neural network is much faster.

Recently, Qu et al. presented a proof-of-concept of application of CNNs for fast reconstruction of high-quality NMR spectra of small, large and disordered proteins from limited experimental data [4a]. Another important result of this work was a demonstration that the highly capable CNN can be successfully trained using solely synthetic NMR data with the exponential functions [28i, 29]. Spectrum aliasing artifacts introduced by NUS data were gradually removed with five consecutive dense CNN blocks with data consistency constrained to the sampled data points [3c] (Figure 6). The experimental result showed that DL can reconstruct high-quality NMR spectra fast. The computational time using the CNN was 4~8% of Low-Rank [28i] for 2D spectra and 12~22% of compressed sensing [28d] for 3D spectra (Figure 7).

An alternative type of the network for reconstructing high-quality protein NMR spectra from NUS data was presented by Hansen [4b]. Unlike dense CNNs which are often used for image processing, in this work, a variant of RNNs, the LSTM network was applied. LSTM networks are traditionally used for time series analysis. Thus, the network reconstructs original FID signals in the time-domain, whereas the CNN [4a] treats the spectra in the frequency-domain as an image. For training of the modified LSTM network, synthetic NMR data was utilized. The input of the network consisted of the NUS time-domain data matrix and a sampling schedule, while the output consisted of the reconstructed NMR data in the time domain and then Fourier transformed it into spectra (Figure 8). The result showed that the intensity of the reconstructed peaks was accurate, albeit the network's computational time was similar to conventional methods for 2D spectra.

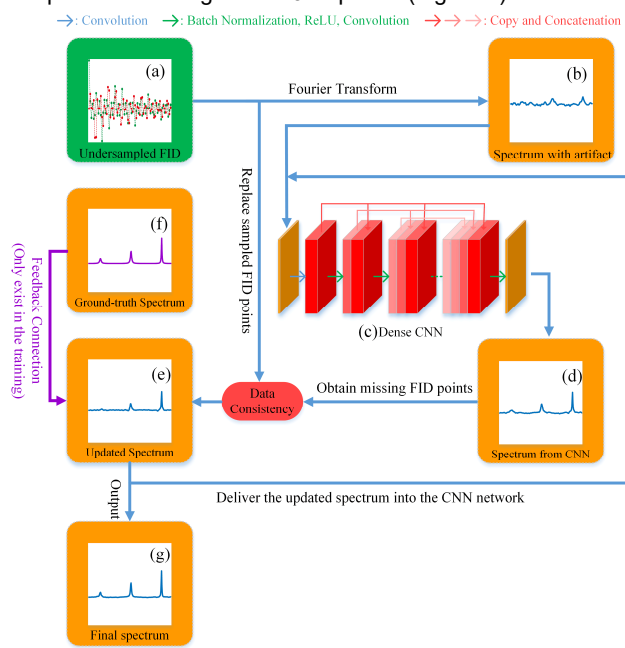

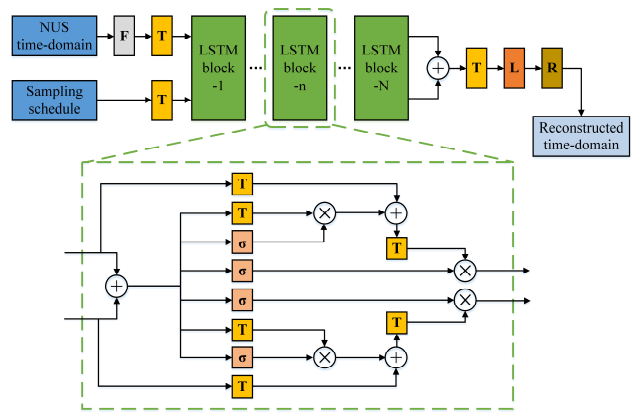

**Figure 8.** The architectures of NMR spectra reconstruction with modified LSTM network. In the green box is the modified LSTM cell. 'F' is the flattening layer, 'T' is the tanh() activation and bias, 'σ' is the sigmoidal activation and bias, '+' is the elementwise addition layer, '×' is the elementwise multiplication layer, 'R' is the reshape layer and 'L' is the linear layer. Adapted from Figure 1 in Ref. [4b].

## 4. Denoising of the Spectra

Relatively low sensitivity of the spectra is a problem that is recurrently addressed in the development of NMR methodology. The *in vivo* brain spectra usually have low SNR and significant overlap between metabolite signals. The problem is exuberated by poor homogeneity of the magnetic field in the studied samples. Denoising is the key process to provide valid information for researchers and physicians [30]. The conventional approach using denoising filters to FID signals in the time-domain is limited by the broad dispersion of decay rates over different spin systems [31]. Furthermore, the signals themselves often generate spectra distortions. For instance, a short Time-of-Echo (TE) signal in macromolecules metabolite may interfere with the target signal by superimposing on the spectral baseline across the entire spectral range.

Although existing denoising filters and J-differential edits [32] can effectively reveal the target metabolite signals from neighboring metabolite signals and distorted spectral baseline, they do not work with all visible magnetic resonance metabolites. Inspired by the robustness of DL, Lee and Kim developed a CNN which was trained and tested on simulated brain spectra with wide ranges of SNR (6.90-20.74) and linewidth (10-20 Hz) [20]. The CNN was further tested on *in vivo* spectra with substantially different SNR from five healthy volunteers (Figure 9). DL clearly managed to infer the mapping between the spectra with lots of interference and the high SNR spectra. Also notable that, similar to the above-mentioned works on spectra reconstruction from NUS data [4], simulation data were successfully used for the network training. The robust performance of the proposed method for low SNR may allow acquiring of a sub-minute $^1$H spectra of the human brain, which would be an important technical achievement for clinical applications.

**Figure 6.** The architectures of NMR spectra reconstruction with CNN. (a) The undersampled FID, (b) the spectrum with strong artifacts, (c) dense CNN, (d) the output of dense CNN, (e) the updated spectrum from data consistency, (f) fully sampled spectrum, (g) the reconstructed final spectrum. Adapted from Figure S1-1 in Ref. [4a].

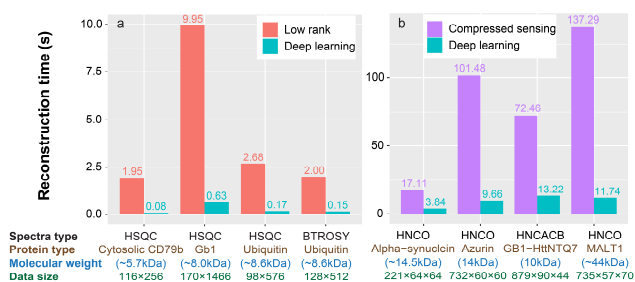

**Figure 7.** Computational time for the reconstructions of (a) 2D spectra and (b) 3D spectra with Low Rank, Compressed sensing and CNN. Note: The listed below each bar are spectra type, the corresponding protein, molecular weight, and spectra size, respectively. For 2D (3D) spectra size of the directly detected dimension is followed by size(s) of the indirect dimension(s). Adapted from Figure 5 in Ref. [4a].

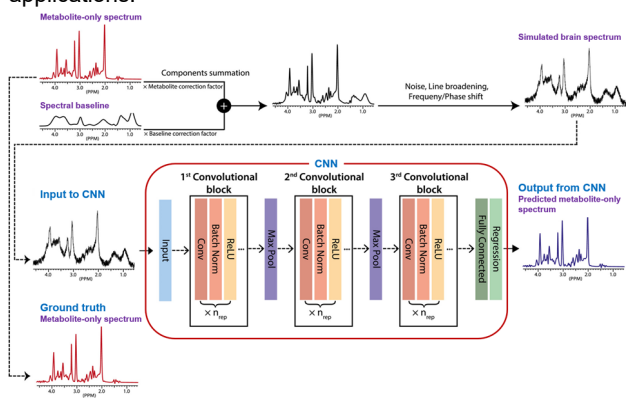

**Figure 9.** The schematic of the simulation of brain spectra and the training of the CNN for denoising. Combined with noises, line broadening, frequency/phase shift and spectral baseline, the metabolite-only simulation spectra can mimic the in vivo brain spectra, which are used as the CNN's input in the training. The network is trained for mapping the brain spectra with such many interference like frequency/phase shift, and unknown baseline into the noise-free to high SNR metabolite spectra. Adapted from Figure 1 in Ref. [20].

## 5. Interpretation of the Spectra

### 5.1. Chemical Shift Prediction

Chemical shift is the most informative parameter obtained from NMR spectra. It is closely related to structural information of compounds, e.g. backbone and side-chain conformation, and can be used to derive 3D protein structure [33]. However, multiple contributions including H-bonding, local electric fields, ring-current effects, etc., make it difficult to use deterministic approaches for calculating the chemical shift values.

The basic problem of obtaining secondary and tertiary structural information of a compound is how to define the complex nonlinear mapping between chemical shift values and structure. Fortunately, the database-derived empirical optimization methods, e.g. ShiftX [34], CASPER [52], PPM [53], SPARTA [35], and Camshift [36], give us great inspiration for learning statistical rules through the training enormous data. DL is a promising approach in this field. DL was used to create relationships between the environmental and structural information of compounds and their chemical shifts.

An early network-based method that predicted the chemical shift from protein structure was PROSHIFT [37]. In 2010, Shen and Bax utilized DNN in SPARTA+ [21] for chemical shift prediction of backbone and $^{13}C^\beta$ atoms, which was trained by an approximately two-fold larger protein database developed for TALOS+ [33f]. In the DNN, the input layer had 113 neurons giving similarity scores of 20 amino acid types for each residue, and each node in the hidden layer received the weighted sum of input layer nodes as an input. The output was obtained through a nonlinear function. SPARTA+ demonstrated consistent although modest improvement (2~10%) over the best methods, and apparently approached the limits of empirical methods for predicting chemical shift. After the success of the SPARTA+, in 2015, Li and Brüschweiler designed the PPM_One [51] with a new DNN for chemical shift prediction. Specifically, the input layer included 113 nodes for non-proton atoms and 122 nodes for protons, and the hidden layer included 25 neurons. Notably, the transfer function from the hidden layer to the output layer was linear which was different from SPARTA+. The performance of PPM_One in chemical shift prediction was better than SPARTA+ for all atoms except the C' carbonyl carbons.

In addition to using the statically structural information of compounds as DNN inputs, Liu et al. [24] attempt to predict the chemical shift using an atom-centered Gaussian density model with DL in 2019. In the model, the evaluated atom is placed at the center of the 3D grid, and its chemical environment is represented by calculation of the density in different grid sizes. Liu et al. designed a Multi-Resolution 3D-DenseNet (MR-3D-DenseNet) which used the evaluated atom's chemical environment as the input. The network mainly consisted of the multiple channels that were utilized for cropping, pooling, and concatenation to define different spatial resolutions for each atom type described by its atom-centered Gaussian density (Figure 10), and predicted the chemical shift by the full connected layer in the end. Take advantage of this dense network, the data flow penetration feature maintained low and high resolution features across the deep layers (Figure 11). The experiment showed a great agreement for $^{13}$C, $^{15}$N, and $^{17}$O chemical shift, and the accuracy of $^1$H chemical shift was highest and comparable to the ab initio quantum chemistry methods.

DL can also address the inverse problem, which is using the chemical shift to predict the compound structure. With the success of SPARTA+, Shen and Bax developed a DNN based TALOS-N for predicting protein secondary structure such as backbone torsion angles from $^1$H, $^{15}$N, $^{13}$C chemical shift [22]. In the first level of the network, the input included six secondary chemical shift values, six chemical shift completeness flag values and twenty amino acid type similarity scores for pentapeptide. Then, the second level fine-tuned the output of the first level and finally predicted the 324-state torsion angle distribution of residue. The validation on an independent set of proteins showed that backbone torsion angles can be predicted from the DNN for a larger, ⩾90% fraction of the residues, with an error rate smaller than 3.5%.

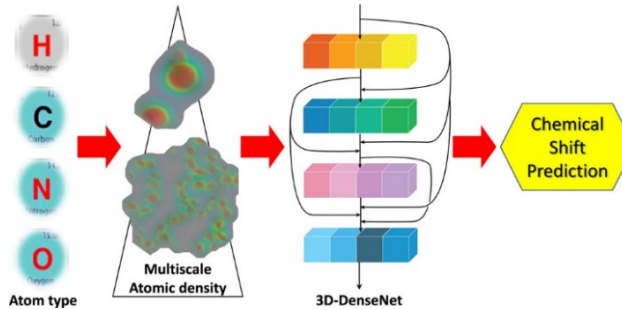

**Figure 10.** The schematic diagram of the overall flow of chemical shift prediction using atomic density. Reproduced from Abstract in Ref. [24]. Copyright 2019, American Chemical Society.

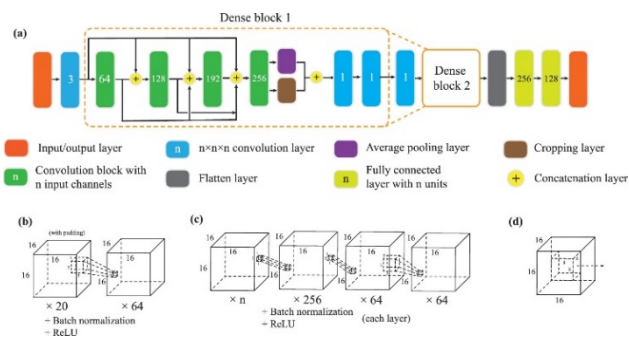

**Figure 11.** The overall architecture of the MR-3D-DenseNet model for chemical shift prediction. (a) The flowchart of the network, (b) the 3×3×3 convolution layer prior to the first dense block, (c) the repeating unit in DenseNet block that contains two 1×1×1 convolution layers followed by a 3×3×3 convolution layer, (d) the cropping layer from the center of the feature map. Reproduced from Figure 2 in Ref. [24]. Copyright 2019, American Chemical Society.

### 5.2. Automated Peak Picking

There is still a large potential of using the artificial intelligence and neural networks in NMR spectroscopy for automation of the laborious data analysis. It usually takes days to months for experienced users to accomplish routine tasks, such as peak picking, resonance assignment, and structure calculation. Automation of the NMR workflow would benefit structural studies of macromolecules, drug discovery, and systems biology. Robust and false-free peak picking is the first and among the biggest challenge for the automation [38]. The main difficulties in automated peak picking come from peak overlap, low SNR, line distortions and presence of spectral artifacts [25]. The first attempt to automate the peak picking dates back to the late 1980s, when most of the approaches utilized features such as symmetry and intensity of the peaks [39]. Subsequently, there were many different automated peak picking methods, mainly including threshold approaches- NMRView [40], XEASY [41], CCPN [42], noise-based methods [43], matrix factorization [44] and Bayesian approaches [45]. Although dozens of peak picking methods are widely available, none of them can fully substitute manual analysis by an expert [45b].

DL has been shown to consistently achieve human-level performance in various recognition tasks, and thus looks like an ideal method for addressing the task of automated NMR detection of signals. Klukowski et al. demonstrated the NMR-Net for peak picking [5]. The method includes the following steps: (1) determine the candidates of the targeted peak by the local extremum in the N-dimensional spectrum. (2) eliminate the candidates whose intensities are low (below noise level). (3) normalize the resolution and intensity of the spectra. (4) classify the peaks. Each peak candidate is fed to a CNN, which returns a real value between 0 and 1, representing the probability of the peak. The overall architecture of the model (Figure 12) consisted of two convolutional layers with max-pooling and the fully connected layer with a sigmoid function. The model input was a matrix of 48×48 pixel values, representing a cropped fragment of the normalized spectrum. This CNN model was verified on 31 manually annotated spectra, and a high top-tier average precision (0.9596, 0.9058 and 0.8271 for backbone, side-chain, and NOESY peaks respectively) was obtained.

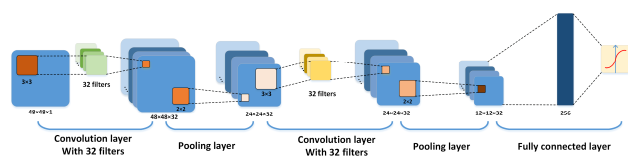

**Figure 12.** The architecture of NMR-Net for automated peak picking. The model feeds the 48×48 2D patches as inputs propagating forward and its outputs are corresponding to the probability of true peaks in NMR spectra. Numbers beside signify the size of the image after processing on the corresponding stage. The final layer, which is the fully connected layer consisting of single neuron has the function of sigmoid activation, to achieve the purpose of classification. Adapted from Figure 2 in Ref. [5].

Another example of the use of DL for peak picking was presented by Bruker Biospin Corporation [26]. Inputs of the network were simulated spectra with labels. During the training phase, the network parameters have been updated in an iterative way, so that the DNN prediction of the simulated training spectra is closer to the corresponding labels with every step. After that, DNN can be used to predict labels on real data. The result on experimental data showed that the trained DNN can accurately define regions corresponding to actual 1D $^1$H NMR signal with an accuracy consistent with the manually selected signal regions.

## 6. Summary and Outlook

Admittedly, DL uses a unique data-feed approach to find complex nonlinear mappings between inputs and outputs. So far, (1) DL successfully helped us to discover the relationship between NMR spectra with noisy and distorted signals and intact spectra. (2) DL replaced complex calculations and manual analysis, such as chemical shift prediction and peak picking. Nevertheless, DL has long been criticized for its lack of interpretability, and it is difficult to understand what the network had learned while implementing various mappings. Recently, Bengio et al. proposed a meta-learning causal structure [46] and Amey et al. presented a group-theoretical procedure [47], trying to open the black-box. Last but not least, the shortage of the training data hinders the development of DL. Many researchers try to solve this problem by building up their training set with the simulated data [4a, 4b, 5, 20, 26] and utilizing large trustworthy databases [21, 22, 24, 51]. Therefore, in the future, it is necessary to establish larger and more diverse databases, to give researchers easy access to different types of NMR data. Meanwhile, training the network using combination of simulated data and experimental data is another important way. BioMagResBank and the Protein Data Bank provide over 11,900 entries containing $^1$H, $^{13}$C, $^{15}$N and $^{31}$P assigned chemical shifts and coupling constants of peptides, proteins and nucleic acids, while NMRbox [54] sets a good example to connect NMR researchers together, which is not only sharing their data set (experimental or simulated), but also data processing tools and programs, to make researchers more convenient for preparing their training data.

With the future development of DL, we may anticipate that more problems in NMR spectroscopy will be addressed and solved. An incomplete list of possible applications may include: (1) the accelerated high-quality reconstruction of high-dimensional biochemical NMR spectra will become possible with the exploration of DL architectures and optimization algorithms. (2) in denoising, removal of residual water signals and other spectroscopic artifacts, which complicate detection and accurate quantification of metabolites, will be considered. (3) in the interpretation, DL may solve complex tasks from chemical shift assignment to the discovery of structures and description of the physical-chemical properties of new compounds. (4) extending to diffusion spectra, dynamic spectra, and large-scale spectral data training, and integrating the time and frequency domain together as the input for more information.

## Acknowledgements

The authors thank Khan Afsar for polishing the paper. The authors are also grateful to researchers for insightful discussions and publishers for adopting figures. This work was supported in part by the National Natural Science Foundation of China (NSFC) under grants 61971361, 61871341, 61571380 and U1632274, the Joint NSFC-Swedish Foundation for International Cooperation in Research and Higher Education (STINT) under grant 61811530021, the National Key R&D Program of China under grant 2017YFC0108703, the Natural Science Foundation of Fujian Province of China under grant 2018J06018, the Fundamental Research Funds for the Central Universities under grant 20720180056, the Science and Technology Program of Xiamen under grant 3502Z20183053, the Swedish Research Council under grant 2015–04614 and the Swedish Foundation for Strategic Research under grant ITM17-0218.